\documentclass[aps,a4,twocolumn,showpacs]{revtex4}
\usepackage{graphics, epsfig, hyperref,subfigure,color}
\begin{document}
\title{Field dependent Anisotropic Micro Rheological and Microstructrual properties of Ferrofluids}
\author{Balaji Yendeti, G. Thirupathi, Ashok Vudaygiri\footnote{corresponding author: email avsp@uohyd.ernet.in}, R. Singh}
\affiliation{School of Physics, University of Hyderabad, Hyderabad, 500046 {\bf India}}
\pacs{66.20.-d, 83.80.Gv, 47.57.-s}
\begin{abstract}
We have measured micro-rheological and micro-structural properties of a super paramagnetic ferrofluid made of  Mn$_{0.75}$Zn$_{0.25}$Fe$_2$O$_4$ (MZF) nanoparticles, using passive microrheology in a home built inverted microscope.  Thermal motion of a probe microsphere was measured for different values of an applied external magnetic field and analysed.  The analysis shows anisotropy in  magnetoviscous effect. Additional microrheological properties, such as storage modulus and loss modulus and their transition is seen. Following the analysis given in Oliver Muller et.al., J. Phys. Cond. Matter {\bf 18}, S2623, (2006) and Stefan Mahle et.al., Phys. Rev. E. {\bf 77}, 016305 (2008) we have obtained microstructural properties such as elongational flow coefficient $\lambda_2$, relaxation time constant $\tau$, coefficient of dissipative magnetization $\alpha$ etc.  From the measured viscosity data, all the above parameters could be obtained. Our values for the above parameters are in agreement with earlier theoretical calculations and macro-rheological experimental measurements.

\end{abstract}  
\maketitle
\section{introduction}
Magnetic fluid  contains  colloidal ferromagnetic nanoparticles in carrier liquids like water, kerosene, Toluene etc. In presence of an external magnetic field, individual particles align to form long chains of nano particles along the direction of the applied field \cite{A.Mertelj}. This modifies the fluid properties of the sample. High enough fields can even achieve a phase transformation, showing an almost  solid-like behaviour.  Furthermore, the system is anisotropic and values for the parameters along the field are different from those values for perpendicular to the field. The extent of this anisotropy can be measured by  passive microrheology. The underlying physics behind  such a behaviour can be better understood by  studying the microrheological properties of these magnetic fluids.  

The chain formation typically happens when the dipole moment induced due to the external field exceeds the average thermal energy $k_BT$ of the nanoparticles. At high fields, the chains are typical 10$-$20 $\mu$m long and are easily visible under microscope. The external field also causes a torque on each particle given by $\mu \textbf{M} \times \textbf{H}$, which constraints their rotation about axis perpendicular to the magnetic field. This causes an additional friction, and hence an increased viscosity \cite{Mc.Tague1969}. At higher fields, all particles are effectively pinned against rotation, which leads to an anisotropy.

 Magnetic fluids have several interesting engineering, bio-medical and scientific applications \cite{C.Scherer, Tobias Neuberger, marcela gonzales, sunghyn yoon}. Hence it has been a subject of several studies earlier, which attempt to understand the diffusive characteristics of ferrofluids.  Magnetic field dependent viscosities, referred as Magneto-viscous effect (MVE), have been studied  in macroscopic regime theoretically \cite{M.Shliomis, Stefan Mahle} as well as experimentally\cite{Mc.Tague1969, S.Odenbach-1}.  Mertelj et. al. have recently reported about 4 times increase of the storage modulus along the external magnetic field in maghemite samples, although their measurements for anisotropy in  MVE were inconclusive due to a larger statistical variation \cite{A.Mertelj}. But earlier theoretical predictions, such as by Morozov\cite{Morozov}, and following capillary viscometer experiments \cite{Mc.Tague1969, S.Odenbach-1} showed a clear anisotropy. Odenbach et. al, explained the anisotropy in  MVE as due to structures formed along the field.  When flow was perpendicular to the applied field, these structures offered stronger hindrance, resulting in higher viscosity \cite{S.Odenbach-1}.

We present here measurements of microrheological properties of a Mn$_{0.75}$Zn$_{0.25}$Fe$_2$O$_4$ (MZF) ferrofluid by recording the  thermal motion of a silica microparticle suspended in it. MZF is a soft ferrite and its nanoparticles are well useful for ferro fluids. Bulk magneto-viscous measurements show a transition from non-Newtonian to Newtonian as shear rates are increased from 0.1 to 1000 s$^{-1}$ at different magnetic fields. Magneto-viscosity exhibits a hysteresis at constant shear rates\cite{thirupati3}.

 Mn-Zn ferrofluids are applicable in magnetic valve controls, fluid transformers, shock absorbers, hypothermia etc., because of its suitable magneto-rheological and electro-rheological properties \cite{marcela gonzales, sunghyn yoon}. Following standard approach of passive micro-rheology, thermal motion of a silica bead was recorded in a home-built inverted microscope and analysed using video tracking software. The trajectories of the probe particle thus measured yields the local probe response and the linear visco elasticity. From these measured values of viscosity, we derive three other relevant parameters by following relationships given in earlier works - viz., as elongational flow coefficient $\lambda_2$ \cite{Oliver Muller, S.Odenbach-3}, coefficient of dissipative magnetization $\alpha$ \cite{Stefan Mahle}, magnetization relaxation time constant $\tau$ \cite{S.Odenbach}. The experimental details and results are described below. 

\section{The Experiment}

\begin{figure}[h]
\subfigure{\includegraphics[scale=0.23]{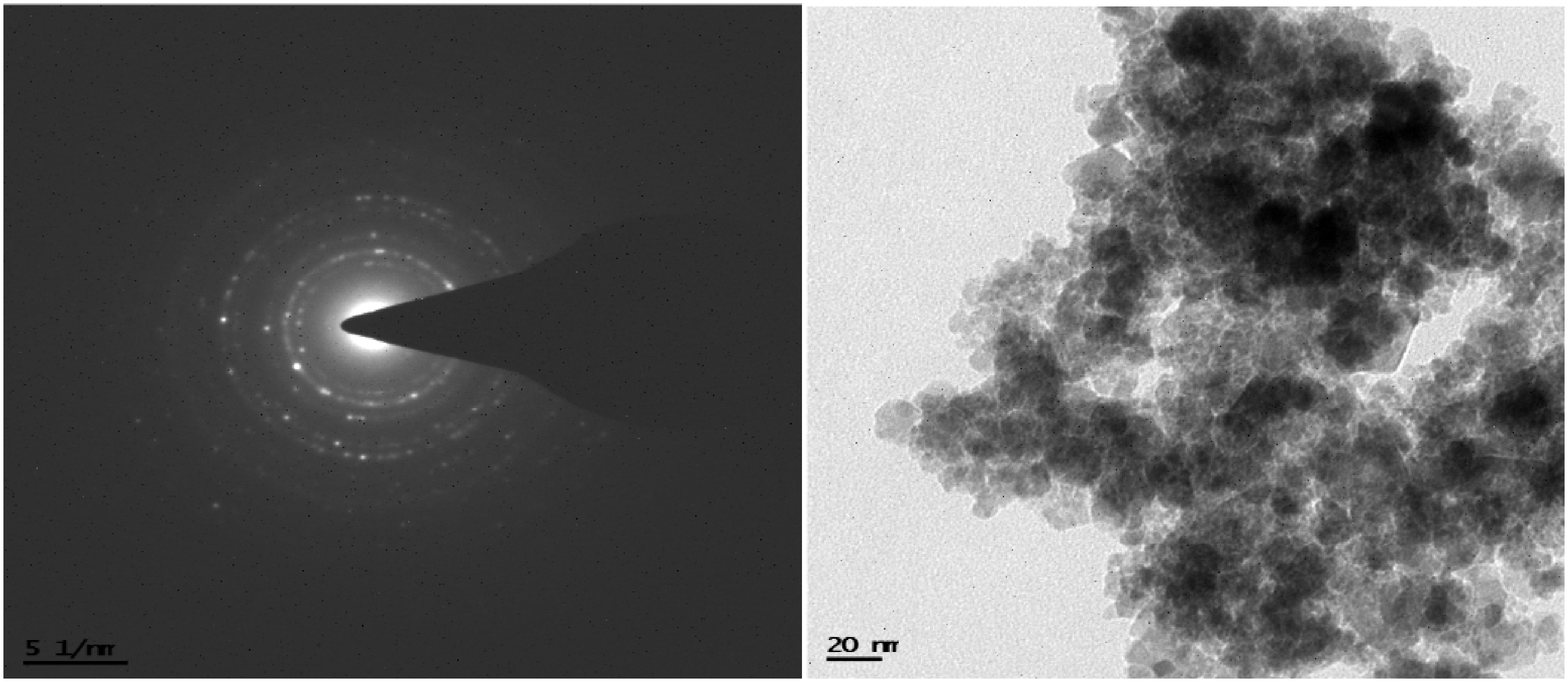}}
\caption{TEM images of nanoparticle.}
\label{tem}
\vskip0.2cm
\subfigure{\includegraphics[scale=0.3]{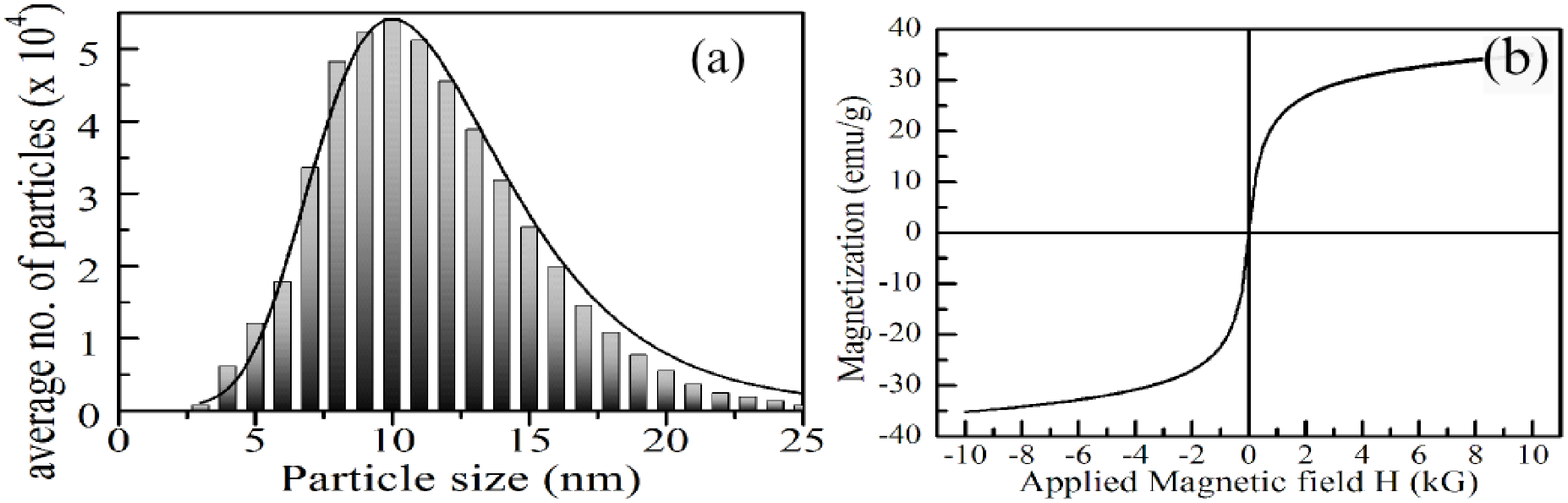}}
\caption{(a) Histogram for particle size distribution  and (b) M-H curve for the sample at room temperature, showing superparamagnetic nature with zero coercivity (right)}
\label{MHcurve}
\end{figure}

Soft ferrimagnetic Mn$_{0.75}$Zn$_{0.25}$Fe$_2$O$_4$ ferro fluid was synthesized by a soft chemical approach of coprecipitation \cite{thirupati1,thirupati2}. The size distribution of nanoparticles was analysed using TEM micro graphs. The electron diffraction pattern shows single phase of the system of nanoparticles (see figure \ref{tem}. The uniformly dispersed Mn-Zn ferrite nanoparticles in Ethylene Glycol (EG) were used for micro rheological measurements.  Magnetization measurements of the sample shows a saturation at as low as 0.24 T and a superparamagnetic nature with zero coercivity and zero remnance - as seen in figure \ref{MHcurve} (b)

From the above sample, 50 $\mu$l of this magnetic nano particle solution, 1 ml of water and 20 $\mu$l of N,N-Dimethyl-N-Octadecyl-3-aminopropyltrimethoxysilyly chloride(DMOAP)  are mixed thoroughly by vortexing for 20 minutes to get uniform alignment of octadecyl chains on the surface of these magnetic nano particles. This addition of DMOAP is for steric stabilization of the nanoparticles. On addition of DMOAP,  one would observe a foam layer on the top of the solution, because of the silane. This mixture is further centrifuged to remove excess solution from nano particles and  1 ml of water is added to this nano particles to wash out the remained silane,then again mixed, centrifuged to  remove the excess solution with nano particle aggregates. This process is done 3$-$4 times until there is no foam layer. Samples with different concentrations, ranging from 4\% to 16\% ratios by weight of nanoparticles were prepared from this stock, by adding different amounts of distilled water to this stock solution. To each of these sample solutions,  2 to 5 $\mu$l of suspension of silica beads (average diameter 2.34 $\mu$m) in water and 30 to 50 $\mu$l of Polyethylene glycol(PEG) solution is added. These samples were taken into a glass cell, made by microscope coverslips sealed on all sides with double sided tapes of thickness 100 $\mu$m. This sample cell is placed in a home built inversion microscope with 100X, 1.4 N.A., oil immersion microscope objective (Olympus, UPLSAPO series). A homogenous magnetic field with  electromagnets is arranged in such a way magnetic field is in the parallel direction with the sample cell and with a gap between the two coils being 20 mm. The gradience in field with in this gap is  less than 1 mT at a magnetic field strength of 240 mT.
		
The thermal motion of the silica bead in this sample is recorded using a normal CCD camera (Watec 90, 512 $\times$ 512 pixels) and a frame grabber card (National Instruments IMAQ NIPCI-1209), at a rate 30 frames/sec acquisition speed and analyzed. A freely available particle tracking software with facility of  auto tracking the bead in a video frame in which a template and its associated target are defined is used which becomes useful in analysing the highly concentrated nanoparticle solution \cite{tracking}

The position of the silica beads in the magnetic nano particle solution was determined by this off-line analysis with an accuracy of $\pm$ 55 nm. The probability distribution of the thermal motion of the particles were fitted with a Gaussian curve. The probability that the silica bead would diffuse to a certain distance $\delta x$ in this magnetic fluid in the time interval `t' is $P=P_0 \exp(-(\delta x)^2/\Delta^2 t)$. Here P$_{0}$ is the normalization constant and $\Delta t$ is the width of the Gaussian distribution (see figure \ref{guassian_histogram}). From this, the diffusion coefficients of the silica bead are determined in the parallel and perpendicular directions of the application of the field with $D_{\parallel,\perp} = \Delta_{\parallel,\perp}^2(t)/4t $ \cite{dhara}. The corresponding parallel and perpendicular viscosities of the sample are measured using the Stokes-Einstein relation 

\begin{equation}
\eta_{(\parallel,\perp)}=\frac{k_{B}T}{6\pi D_{(\parallel,\perp)}r},
\end{equation}
where  `r' is radius of the micro bead.

For these measurements, four different sample cells with different concentrations ranging from 16\% ratio by weight to 4\% ratio  by weight were used. For each cell, measurements were obtained for six different magnetic field values ranging from zero to 240 mT. At each field value six different beads are tracked for approximately 3 minutes each at a rate of 30 frames per second. The statistical variation of measurement arising due to this is shown as error bars. 

Optimization of the measurement was verified by measuring the viscosity of 40\% glycerol and 60\% water.  All error bars in these measurements are statistical. All other micro-rheological parameters of the ferrofluid are derived from these data of thermal motion, as explained in the following section. 

\section{Results and discussion}
\subsection{Magnetic properties}

Super paramagnetic particles under the action of an external magnetic field, acquire induced magnetic dipoles and these dipoles interact to form long chains. This interaction energy between these magnetic dipoles is given by \cite{Jordi Faraudo},
				
\begin{equation}
U=\frac{\mu_{0} m^{2}}{4 \pi r^{3}}[1-3\cos^{2}\theta]
\label{interaction}
\end{equation}			
Where $\mu_{0}=4\pi \times 10^{-7}N/A^{2}$ is the magnetic permeability of free space, $m$ is the magnetic dipole moment, $r$ is the distance between the two magnetic dipoles and $\theta$ is the angle between the applied external magnetic field and the line joining between the  centres of the two magnetic dipoles. Since, this dipole-dipole interaction energy is dependent on $\theta$ value, anisotropic structures with long chain formations are formed in these super paramagnetic fluids. This interaction is dominant over a  characteristic length scale $\lambda_B$, in comparison to length scale of thermal effects, given by  
\begin{equation}
\lambda_{B}=\left( \frac{\mu_{0}m^{2}}{2\pi k_{B}T}\right) ^{\frac{1}{3}}.
\end{equation} 
The above equation comes from the ratio between the dipole-dipole interaction energy $U$ and thermal energy k$_{B}$T. The minimum energy configuration is when $\theta$=0,  and when the interparticle distance is almost equal to the diameter of the particles `r' the interaction energy equation (\ref{interaction}) becomes 
\begin{equation}
U_{\rm min} =\frac{\mu_{0}m^{2}}{2\pi d^{3} k_{B}T} 
\end{equation}
 
           where `$d$' is the distance between the centres of magnetic nano particles. From the above equation it is very clear that, the field induces a magnetic moment in the  nanoparticles and hence influences the interaction parameter between the magnetic nano particles. In turn, this influences the rheological properties. 
           
\subsection{Micro-rheological properties}

\begin{figure}[h]
\includegraphics[scale=0.3]{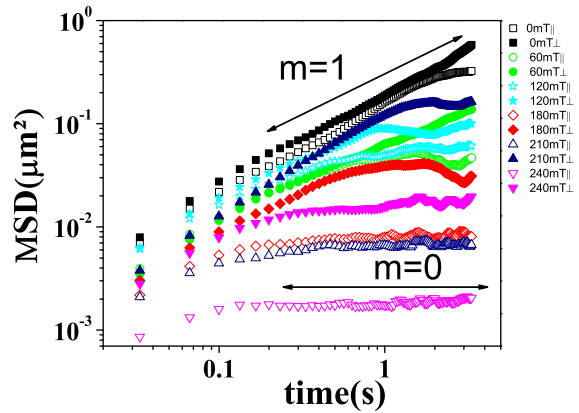}
\caption{(Color online) Mean square displacements of probe particle in magneto rheological fluid for different external magnetic field strengths. Parallel to the external field  (empty) and perpendicular to external field (filled) components are plotted as a function of time(T). Nanoparticles form chains along the magnetic field direction. Suggested slope for the curves are represented as m=1 (zero field) and m=0 (high field)}
\label{msd}
\end{figure}

 From the thermal trajectories of the probe particles, the time dependence of the mean square displacements(MSDs) in the direction along and perpendicular to the external field and are as shown in (Fig.\ref{msd}). As the field strength increases, there is a clear transformation in the slope of these curves. This transformation in the slope represents the transformation\cite{Denis Wirtz} from viscous to viscoelastic and then on to solid behaviour in the chain formation as the magnetic field increases. This is the typical behaviour in repeated measurements. Also, there is increased differences in the MSDs of parallel and perpendicular directions as the field increases. Slope of parallel direction MSDs becomes zero with the increase in field than the slope of perpendicular direction MSDs.

\begin{figure}[!h]
\centerline{\includegraphics[width=7cm]{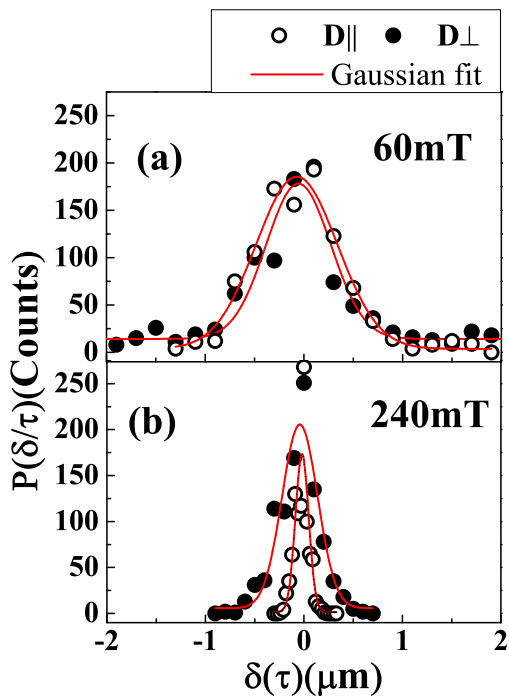}}
\caption{(Histogram of particle displacements parallel (open circles) and perpendicular (closed circles) to the magnetic field for $\tau$ = 1 s. The solid lines are Gaussian fits. Top figure (a) is for low magnetic field (60 mT). Bottom one (b) is for is for high field (240 mT). Anisotropy between parallel and perpendicular behaviour is clearly visible for high field case. }
\label{guassian_histogram}
\end{figure}						 
						 
Histograms for thermal trajectories of the bead  and a curve fit using a Gaussian curve to them is shown in figure \ref{guassian_histogram}. Both parallel component (circles) and perpendicular component (closed circles) are shown for low magnetic field (60 mT - Figure \ref{guassian_histogram} top ) and for a high field (180 mT - Figure \ref{guassian_histogram} bottom). It can be seen that the parallel and perpendicular components almost overlap for zero field situation, while the the parallel component has a significantly narrow distribution compared to the perpendicular component in presence of a high field. This indicates that the corresponding diffusion coefficients follow the relation $D_{\perp}>D_{\parallel}$, and the silica bead finds it easier to move in a direction perpendicular to the applied field than in parallel direction.

\begin{figure}[h]
\includegraphics[scale=0.32]{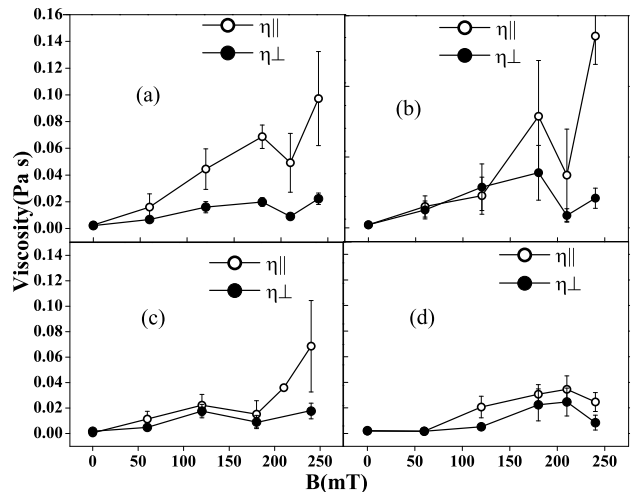}
\caption{Field dependent variation of $\eta_{\parallel,\perp}$ with varying concentration of superparamagnetic particles. Concentrations are for 16\%, 14\%, 10\% and 8\% by weight, respectively for (a), (b), (c), (d)}
\label{magnetoviscous}
\end{figure}					
					
Figure \ref{magnetoviscous} shows parallel and perpendicular viscosities as a function of field, with subfigures (a) - (d) are for different concentrations, as explained in the associated caption. As the field strength increases, the interaction between magnetic nanoparticles dominates the thermal motion of the individual nanoparticles and clusters and they start forming chains. With this,  parallel viscosity increases with increasing field. Perpendicular component of the viscosity also increases, but much more slowly than the parallel component. At very dilute situations such as shown in figure \ref{magnetoviscous} (d), the difference between parallel and perpendicular components are very small. But at higher concentrations, the difference between the two components become much larger, at higher fields. 

Though there will be very little flow in the sample cell, the thermal motion of the solvent brings a small movement in the magnetic chains which leads to torque acting on the magnetic nanoparticle chains. However, there is an important discussion which is relevant with respect to the distribution of the silica beads among the nanoparticle chains, as given below.  
										
\begin{figure}[h]
\includegraphics[scale=0.24]{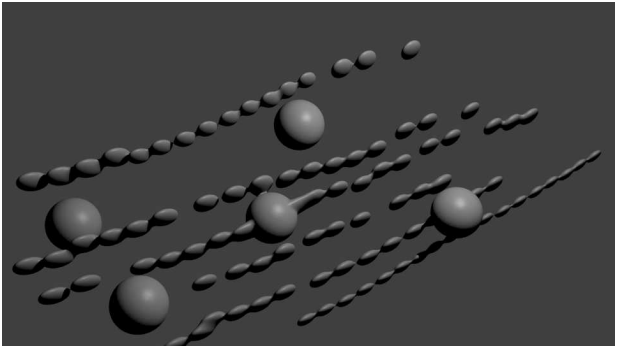}
\caption{A representative rendering of chain formation by magnetic nanoparticles (irregular particles) and the position of silica beads (regular spheres) in relationship to them. Cases mentioned in text as (i), (ii), (iii) and (iv) are clearly seen in figure}
\label{3dfigure}
\end{figure}

We can distinguish four situations of position of the bead with respect to the chains. (i) During the chain formation, some of these probing beads get caught between two closely lying chains and its movement would be restricted to the region within.  (ii) The Silica beads, which are much larger in diameter in comparison to thickness of the chains, sit astride more than one chain. In this case, the bead moves with the chains much like a person sitting on a hammock. The nanoparticles within the chains experience a torque due to the external field and also interaction between each other. This leads to a collective movement of the chain as a whole and this movement is more in direction perpendicular to the chains than along it. This therefore induces a movement to the bead as well and that results in an anomalously high diffusion coefficient in perpendicular direction. (iii) The third case is a situation where the chains are sparse, at least in some regions and a silica bead is almost touching only one chain. In this case, the beads moves easier in the parallel direction, along the chain, giving a higher value for the parallel component. (iv) Fourth case is when the bead is away from any of the chains and gives a true thermal motion. But in higher magnetic field, this region would have very few nanoparticles since most of them are involved in chain formation and thus the viscosity measured in this region would be mostly that of carrier liquid rather than that of ferrofluid. All these situations are schematically represented in the figure \ref{3dfigure}.  For better measurement, we measured thermal motion of different silica beads, spread over all the above situations and then computed their average value. The values shown in figure \ref{magnetoviscous} are a result of such a measurement and the statistical variation between these values will obviously be large. This is the reason for such large error bars depicted in the figure \ref{magnetoviscous}.

In addition, both components of viscosity show a sharp dip at around 210 mT, as seen in figure \ref{magnetoviscous}. Following the same line of arguments as given in \cite{S.Odenbach-1}, we propose the reason for this dip as follows. When nanoparticles of slightly different sizes are involved, the larger particles become part of the chains at low field strengths. The smaller particles, whose magnetic moments are weaker, require slightly higher fields, so as to be pinned down within the chain. However, due to a lower surface contact, particularly between two nanoparticles of different sizes, the smaller particles are also equally likely to come out of the chain at higher fields and form independent structures. The silica bead has to find pathway between these structures, which would perhaps lead to an increase in perpendicular viscosity but a decrease in parallel viscosity. This situation would lead to a decrease in anisotropy, which is seen at fields beyond 210 mT.

\begin{figure}[h]
\includegraphics[scale=0.32]{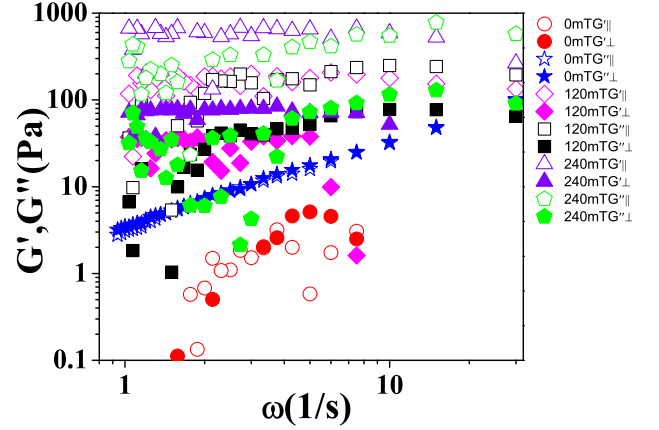}
\caption{(color online)Storage moduli (G$'$) and loss moduli (G$''$)  for different magnetic fields as a function of deformation frequency$(\omega)$. At low fields the system is isotropic as indicated by both parallel components (open symbols) and perpendicular components (closed symbols) have similar values. At higher fields, parallel components have much higher values than perpendicular components, indicating onset of anisotropy. }

\label{all_g}
\end{figure}

From the measured thermal motion of the silica beads, we can also compute  viscoelastic shear moduli(storage modulus$(G')$ and  loss modulus $(G'')$), using generalised Stokes-Einstein equation. G$'$ and G$''$ represent the fraction of energy induced by the  deformation imposed on the material. This fraction of energy is lost by the viscous dissipation in case of loss modulus(G$''$), and, is stored elastically in case of storage modulus(G$'$). G$'$ and G$''$ were computed using the  equations given in Ref.\cite{T.G.Mason-2}, given below,  as a function of rate of deformation $\omega$. They are shown in Figure  \ref{all_g}. 

By expanding the equation of motion of Mean square displacement around $t=1/\omega$, the translational parameter  $\beta$ is defined as \cite{T.G.Mason-2}
									
\begin{equation}
\beta(\omega)=\left. \frac{\langle d \ln \Delta r^2(t)\rangle}{d \ln t} \right|_{t=\frac{1}{\omega}}
\end{equation}

and the Fourier transform of this MSD equation would lead to give G$'$($\omega$) and G$''$($\omega$) as

\begin{eqnarray}
G'(\omega)&=&\vert G^{*}(\omega)\vert \cos\left(\frac{\pi\beta(\omega)}{2}\right) \cr
\cr
G''(\omega)& =& \vert G^{*}(\omega)\vert \sin\left(\frac{\pi\beta(\omega)}{2}\right)
\end{eqnarray}
Where $\vert G^{*}(\omega)\vert=k_{B}T/\pi a \langle \Delta r^2(\frac{1}{\omega})\rangle \Gamma[1+\beta(\omega]$, and `$a$' the radius of the probing bead.

\begin{figure}[h]
\centerline{\includegraphics[width=8.5cm,height=6cm]{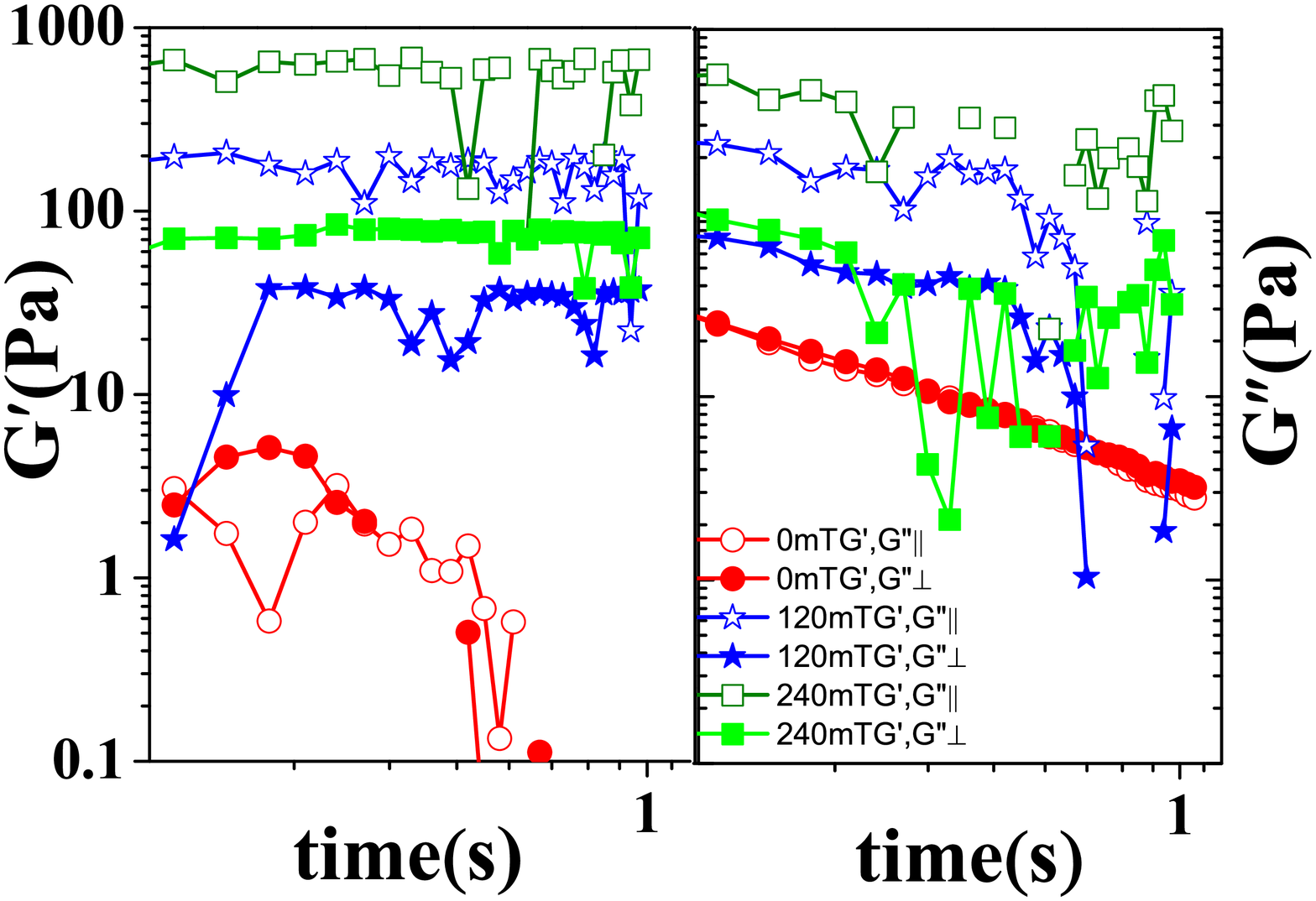}}
\caption{Values of storage (G$'$) (left) and loss (G$''$) (right), moduli for different magnetic fields as a function of time. It can be noticed that G$''$ dominates G$'$ at lower fields. At intermediate fields, their values are closer to each other. At much higher fields G$'$ becomes significantly higher than G$''$. }
\label{gprime}
\end{figure}

From Fig.7 it is indicative that, at very low fields, G$'_\parallel$ and G$'_\perp$ are close to each other (and similarly G$''_\parallel$ and G$''_\perp$) indicating a nearly isotropic behaviour. As field strength increases, the system becomes anisotropic with parallel and perpendicular components becoming different.			
Figure \ref{gprime} shows that, at very low fields, G$''$ dominates G$'$, indicating a pure diffusive movement of the bead, as appropriate of a viscous fluid. Further  At intermediate field strengths,  G$'  \simeq$ G$''$ indicating a viscous to visco-elastic transition. Further increase in field strength results in G$'$ dominating G$''$.  At this stage, G$'$ becomes constant with respect to frequency, which is indicative of a purely elastic behaviour as in case of a solid\cite{Juan}. This Figure \ref{gprime} represents a total transition of the super paramagnetic fluid from viscous nature to the solid characteristics.

\subsection{Micro-structural properties}					
				The other important parameters, such as transport coefficient $\alpha$, the relaxation timescale  $\tau$, and elongational flow coefficient $\lambda_{2}$ were also derived using equations from references \cite{Oliver Muller, Stefan Mahle}. Parallel and perpendicular components of each $\alpha$ and $\tau$ were derived at the limit of small shear. The two components are equivalent to two distinct situations mentioned in references \cite{Oliver Muller, Stefan Mahle}, viz., (a) external field is perpendicular to flow velocity and therefore along the flow gradient and (b) external field is along the flow velocity and therefore perpendicular to the flow gradient.  In our case, the  flow is the small thermal drift of the chains due to external field. 
				
				The measured values of viscosities in the parallel and perpendicular direction to the applied external field are  substituted in the equation \ref{eta_to_eta}, This   gives  $\lambda_{2}$, which accounts for the microstructural properties of ferrofluid with the coupling of symmetric velocity gradient $(v_{ij})$ \cite{definition} to the magnetization relaxation($\tau$) and it accounts for the elongational flow\cite{S.Odenbach-3,S.Odenbach-2}. 
				
\begin{equation}
\frac{\eta_{\parallel}(\xi\rightarrow 0)}{\eta_{\perp}(\xi \rightarrow 0)}=\frac{(1-\lambda_{2})^{2}}{(1+\lambda_{2})^2}(1+\chi)^{2}
\label{eta_to_eta}
\end{equation}
 Where $\chi$ is the magnetic susceptibility.

At higher values of external field, our results show that $\eta_{\parallel}(\xi\rightarrow 0)>\eta_{\perp}(\xi \rightarrow 0)$. This implies that the left side of equation (\ref{eta_to_eta}) is always greater than 1. Rearranging the terms of equation (\ref{eta_to_eta}), this would lead to the result that $\lambda_2>0$. This is also in concurrence with results of Odenbach \cite{S.Odenbach-3, S.Odenbach-2} wherein $\lambda_2$ is shown to be related to the chain length, which has to be a positive value. However, using equation \ref{eta_to_eta}, some of our values for $\lambda_2$ were found to be negative. We did not find any reference to this discrepancy in  the literature till date. However, we propose that an absolute value of $\lambda_2$ be chosen always.  
                               
The reaction of the magnetization to changes in field strength and direction, is relaxation of magnetization. In super para magnetic fluids, the smaller relaxation time will dominate the magnetic relaxation behaviour of the fluid. Here, we show how the motional behaviour or viscosity properties of magnetic nanoparticles in solvents is used to derive slow relaxation times of the magnetic nanoparticles. The relationship between viscosity and magnetization relaxation, under limit of zero shear, is given by \cite{Stefan Mahle} as,

\begin{eqnarray}
\eta_{\parallel} & = & \frac{(1-\lambda_{2})^{2}\tau\chi H_{0}^{2}}{4} \cr
\cr
\eta_\perp & =&  \frac{(1+\lambda_{2})^{2}\tau\chi H_{0}^{2}}{4(1+\chi)^{2}}
\label{stefan_mahle}
\end{eqnarray}

\begin{figure}[h]

\subfigure{\includegraphics[scale=0.35]{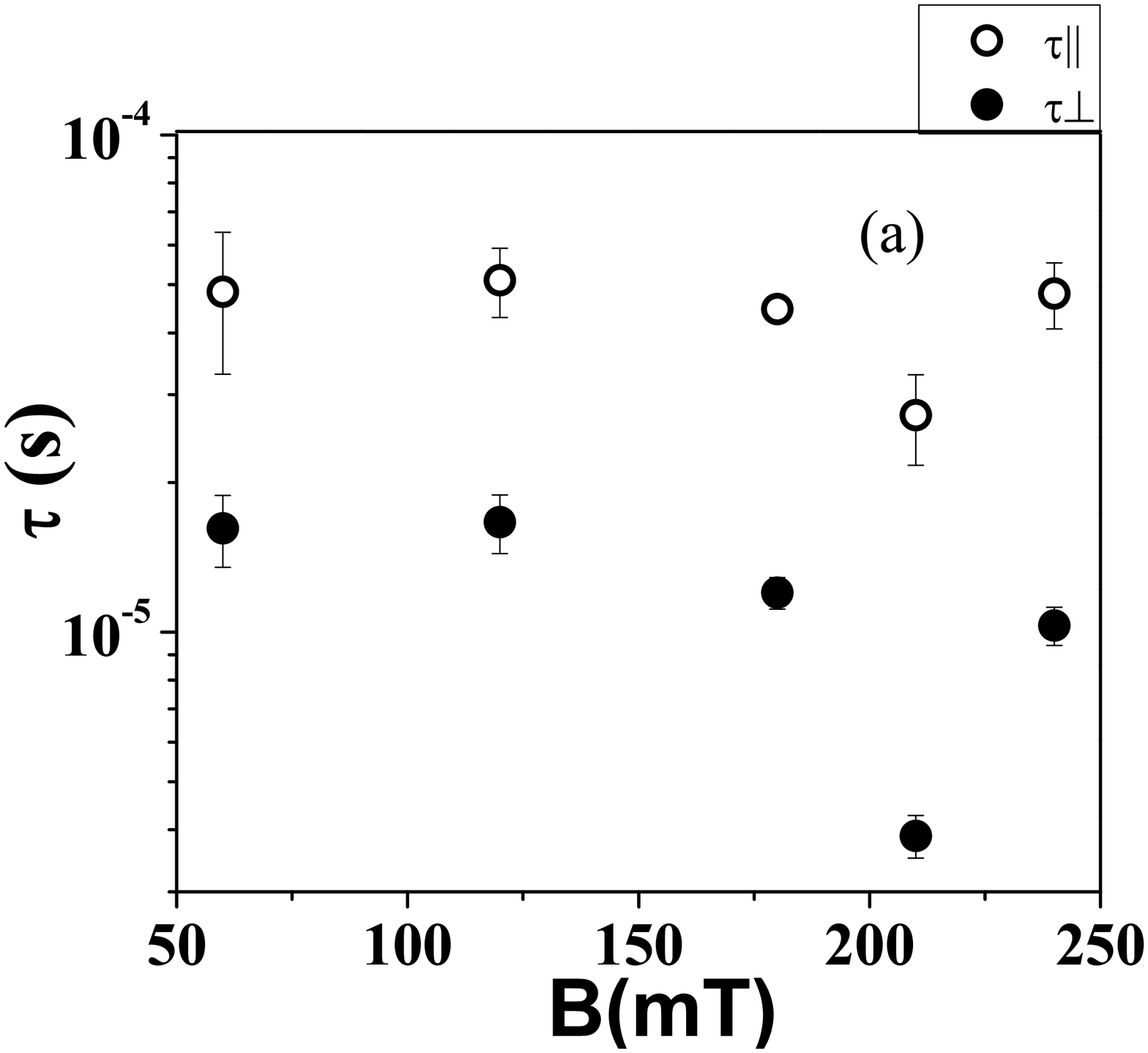}}
\subfigure{\includegraphics[scale=0.35]{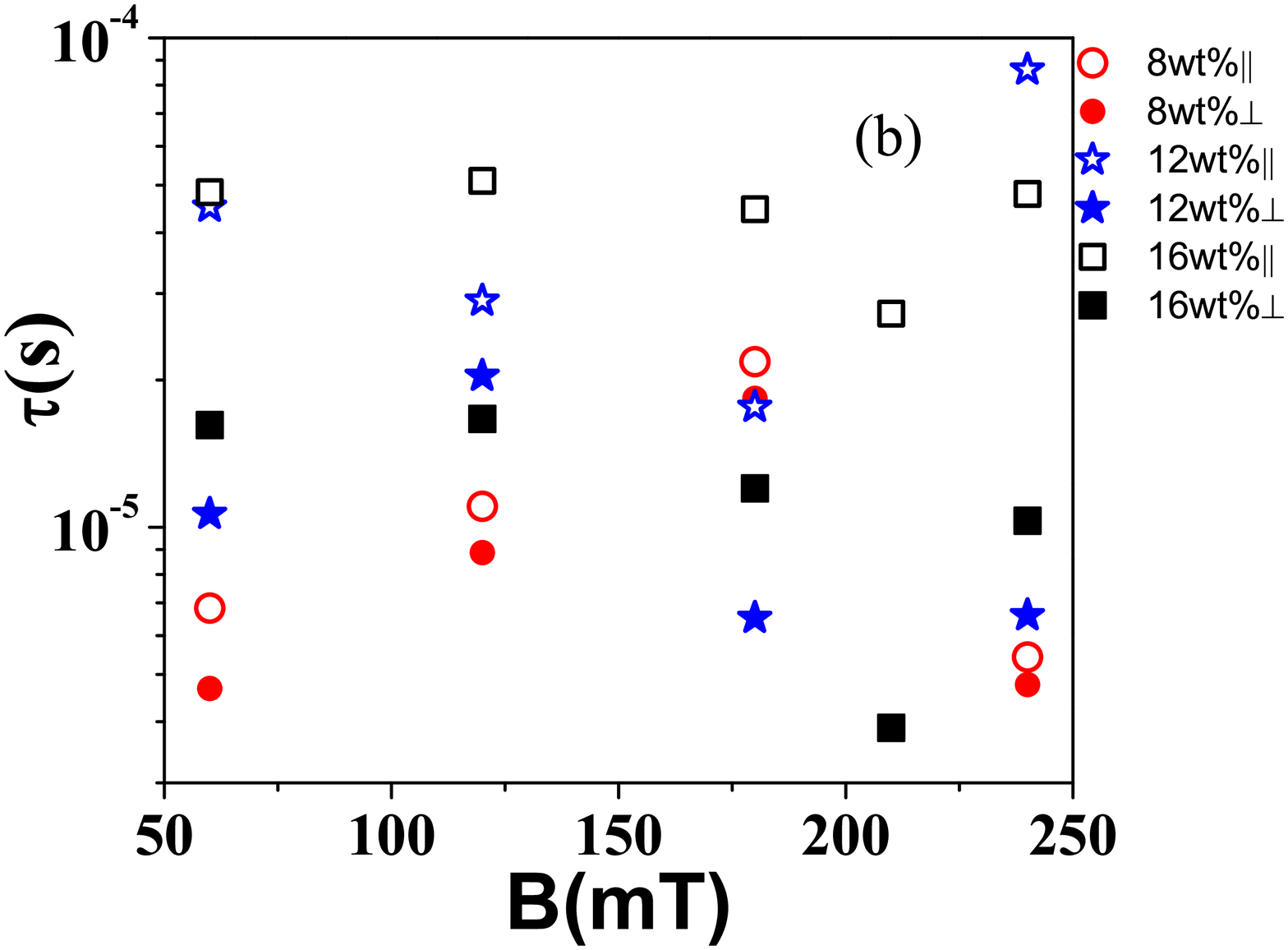}}
\caption{Relaxation time scale $\tau$ as a function of field in (a)(above) and (b) for different concentrations (below). Perpendicular components are indicated by filled symbols and parallel are by open symbols.}
\label{relaxation}
\end{figure}

Relaxation times $\tau$ are obtained by substituting the measured values of $\eta_\parallel$ and $\eta_\perp$  and the derived values of $\lambda_{2},~ \chi$ and H$_{0}$ in the equations \ref{stefan_mahle}, and are plotted in figure \ref{relaxation}. The $\lambda_2$ values derived in our setup ranged from 0 to 0.88 across all concentrations.  While both parallel and perpendicular components of $\tau$ decrease  at higher field strengths, the perpendicular component decreases by more than 10 times the parallel counterpart. It is well known that the relaxation of magnetization consists of two contributions - the Neel relaxation, arising due to change of intrinsic magnetization within the nanoparticles and the Brownian relaxation, which arise due to rotational diffusion of the particles . Below a critical particle size, the value of Neel relaxation are smaller than that of Brownian relaxations \cite{S.Odenbach}. In a polydispersed ferrofluid solution, the smaller particles have much weaker magnetization and therefore faster relaxation times in comparison to larger particles. Also, the relaxation times for smaller particles would be more critically dependent upon the external magnetic field, since the inter-particle interaction is lesser. These would allow us to surmise that the parallel component of $\tau$ is mostly contributed by the Brownian relaxation rate while the perpendicular component is mostly due to the smaller Neelian relaxation and also due to the smaller particles. This would be possible if at lower fields, the larger particles start forming chains at first and the smaller particles join them only at higher fields.

Another important coefficient, $\alpha$ which is related to the dissipative field$(H_{D})$\cite{Stefan Mahle} can be estimated with $\eta=\eta_{0}+\frac{1}{4}\alpha B^{2}$. This $\alpha$ gives the dependence of viscosities of the magnetic fluid. Where  $\eta_{0}$ is the viscosity at zero magnetic field.

\begin{figure}[h]
\includegraphics[scale=0.35]{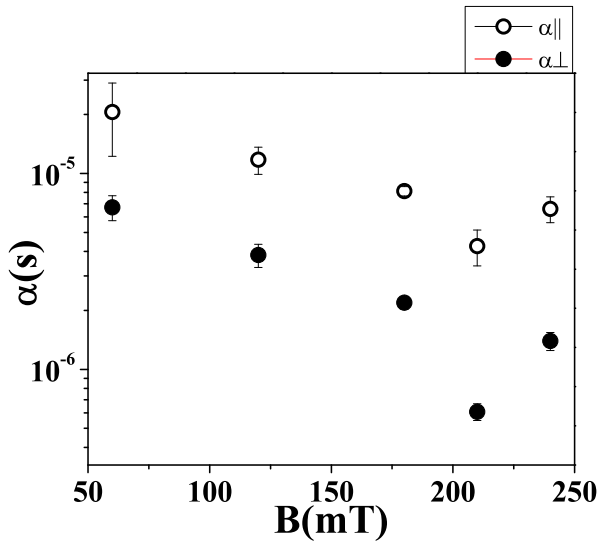}
\caption{Field dependent variation of $\alpha$}
\label{alpha_b}
\end{figure}

Figure \ref{alpha_b} shows that, the $\alpha$ value decreases with the increase in magnetic field strength, which is expected. Because, As the field strength increases, the deviation from the equilibrium magnetization is accounted for by the dissipative field. This dissipative field in turn related to $\alpha$ value using the relation
\begin{equation}
H^{D}=\alpha\left( \frac{\partial}{\partial t}+\upsilon.\nabla-\Omega\times\right)B
\end{equation}

 Also, the difference between parallel and perpendicular components shows that, the dissipation of magnetic field   is lesser in the parallel direction of the applied field compared to the dissipation in the perpendicular direction.   When compared with Shliomi's result, with the equation, $\eta=\eta_{0}+\frac{1}{4}\tau MH$ These values of $\alpha$ are comparable to the values of $\tau$. Which shows the reliability of these values. With the diluted magnetic fluids, one cannot calculate this transport coefficient($\alpha$) value with this formula, because  $\eta\simeq\eta_{0}$.

\section{Conclusion}                                  
We have measured micro-rheological properties of a Manganese Zinc ferrofluid by using thermal motion of a 2.3 micron silica bead suspended in it. The measurements were done using video tracking in a home built inverted microscope.  With this, we are able to measure the magnetoviscous effect, and show a field dependent viscosity. While in most of earlier literature  explains the anisotropy in terms of shear and torque acting on a single nanoparticle (\cite{S.Odenbach-5} for example), we propose a practical consideration of torque on the entire chain, which leads to the anisotropy. 

	In addition,  we are also able to derive, from the statistics of thermal motion, other micro-rheological properties such as the storage modulus $G'$, loss modulus $G''$ (which together form the  complex response modulus $G'+i G''$), elongational flow coefficient $\lambda_2$,  transport coefficient $\alpha$ and relaxation timescale $\tau$. The coupling parameter $\lambda_2$ to the symmetric velocity gradient is calculated from measured viscosity values -  absolute value of this parameter is proposed. The range of these values of $\lambda_2$ are in good agreement with the previously published values from theoretical and macro-rheological methods \cite{S.Odenbach-3}. For the first time, values of $\tau$ are derived adopting theory to experimental micro-rheology values both in parallel and perpendicular directions. We are able to derive all the rheological and structural parameters of the ferrofluid using  measurements of thermal motion of the probe bead. The values of these parameters are very close to those proposed by theory or those obtained by macro rheology measurements.

\section{Acknowledgement}
We thank the Advanced Center for High Energy Materials, University of Hyderabad for financial support, for the microscope setup. Y. Balaji thanks ACRHEM also for the fellowship. We also thank Surajith Dhara for equipment used in preparation of sample cells.

\end{document}